\documentclass[aps,prx, superscriptaddress, twocolumn]{revtex4-1}

\usepackage[a4paper,left=2cm,right=2cm,top=2.5cm,bottom=3cm]{geometry}
\usepackage{graphicx}
\usepackage{amsmath}
\usepackage{amssymb}
\usepackage{color}
\usepackage[colorlinks = false, citecolor=green, linkcolor=red, breaklinks=true, linktocpage=true]{hyperref}
\usepackage{paralist} 

\newcommand{\ket}[1]{|{#1}\rangle}
\newcommand{\bra}[1]{\langle{#1}|}

\newcommand{\tr}{\mathrm{tr}}
\newcommand{\tracesr}{\text{tr}_{\scriptscriptstyle\mathcal{S}\cup \mathcal{R}}[}
\newcommand{\traces}{\text{tr}_{\scriptscriptstyle\mathcal{S}}[}
\newcommand{\tracer}{\text{tr}_{\scriptscriptstyle\mathcal{R}}[} 
\newcommand{\trace}{\text{tr}[}

\newcommand{\lin}{\text{ln}}
\newcommand{\pbeta}{\partial_{\beta}}

\newcommand{\average}[1]{\langle #1 \rangle}

\newcommand{\sr}[1]{{#1}_{\scriptscriptstyle\mathcal{S}\cup \mathcal{R}}} 
\newcommand{\sys}[1]{{#1}_{\scriptscriptstyle\mathcal{S}}}
\newcommand{\res}[1]{{#1}_{\scriptscriptstyle\mathcal{R}}}
\newcommand{\hilbsr}{\mathcal{S}\cup\mathcal{R}}
\newcommand{\hilbs}{\mathcal{S}}
\newcommand{\hilbr}{\mathcal{R}}
\newcommand{\statesr}{\sr{\hat{\pi}}(T)}
\newcommand{\states}{\sys{\hat{\pi}}(T)}

\newcommand{\partsr}{\sr{\mathcal{Z}}}
\newcommand{\partr}{\res{\mathcal{Z}}}
\newcommand{\hamsr}{\sr{\hat{H}}}
\newcommand{\hams}{\sys{\hat{H}}}
\newcommand{\hamr}{\res{\hat{H}}}
\newcommand{\vint}{\sr{\hat{V}}}
\newcommand{\expo}[1]{e^{-\beta#1}}

\newcommand{\hmf}{\sys{\hat{H}}^{*}(T)}
\newcommand{\parts}{\sys{\mathcal{Z}}^*}

\newcommand{\enmf}{\sys{\hat{E}}^*}
\newcommand{\enmfT}{\sys{\hat{E}}^*(T)}
\newcommand{\deriv}{\pbeta\big[\beta\hmf\big]}

\newcommand{\hata}{\hat{A}_\theta}
\newcommand{\hatb}{\hat{B}_\theta}
\newcommand{\hatst}{\hat{\rho}_\theta}
\newcommand{\hatsigt}{\hat{\sigma}_\theta}

\newcommand{\lol}{\Delta}
\newcommand{\hp}{\hat{p}}
\newcommand{\hx}{\hat{x}}
\newcommand{\ha}{\hat{a}_{\scriptscriptstyle T}}

\newcommand{\hA}{A_{\scriptscriptstyle T}}
\newcommand{\hD}{\alpha_{\scriptscriptstyle T}}
\newcommand{\hG}{g_{\scriptscriptstyle T}}
\newcommand{\omt}{\omega_{\scriptscriptstyle T}}
\newcommand{\mt}{M_{\scriptscriptstyle T}}
\newcommand{\bom}{\frac{\beta\omt}{2}}

\begin{document}

\title{Energy-temperature uncertainty relation in quantum thermodynamics}

\author{H. J. D. Miller}
\email{hm419@exeter.ac.uk}
\affiliation{Department of Physics and Astronomy, University of Exeter, Stocker Road, EX4 4QL, United Kingdom.}

\author{J. Anders}
\email{janet@qipc.org}
\affiliation{Department of Physics and Astronomy, University of Exeter, Stocker Road, EX4 4QL, United Kingdom.}

\begin{abstract}

It is known that temperature estimates of macroscopic systems in equilibrium are most precise when their energy fluctuations are large. However, for nanoscale systems deviations from standard thermodynamics arise due to their interactions with the environment. Here we include such interactions and, using quantum estimation theory, derive a generalised thermodynamic uncertainty relation valid for classical and quantum systems at all coupling strengths. We show that the non-commutativity between the system's state and its effective energy operator gives rise to quantum fluctuations that increase the temperature uncertainty. Surprisingly, these additional fluctuations are described by the average Wigner-Yanase-Dyson skew information. We demonstrate that the temperature's signal-to-noise ratio is constrained by the heat capacity plus a dissipative term arising from the non-negligible interactions. These findings shed light on the interplay between classical and non-classical fluctuations in quantum thermodynamics and will inform the design of optimal nanoscale thermometers.

\end{abstract}


\maketitle

\section*{Introduction}

Bohr suggested that there should exist a form of complementarity between temperature and energy in thermodynamics similar to that of position and momentum in quantum theory \cite{Bohr1932}. His reasoning was that in order to assign a definite temperature $T$ to a system it must be brought in contact with a thermal reservoir, in which case the energy $U$ of the system fluctuates due to exchanges with the reservoir. On the other hand, to assign a sharp energy to the system it must be isolated from the reservoir, rendering the system's  temperature $T$ uncertain. Based on this heuristic argument Bohr conjectured the thermodynamic uncertainty relation:
\begin{align}\label{eq:tur}
	  \lol \beta \geq \frac{1}{\lol U},
\end{align}
with $\beta=(k_{\text B} T)^{-1}$ the inverse temperature. While \eqref{eq:tur} has since been derived in various settings \cite{Mandelbrot1956,Phillies1984,Gilmore1985,Lindhard,schlogl1988,Uffink1999,Velazquez2009,Falcioni2011}, it was Mandelbrot who first based the concept of fluctuating temperature on the theory of statistical inference. Concretely, for a thermal system in canonical equilibrium, $\lol \beta$ can be interpreted as the standard deviation associated with estimates of the parameter $\beta$. Mandelbrot proved that~\eqref{eq:tur} sets the ultimate limit on simultaneous estimates of energy and temperature in classical statistical physics \cite{Mandelbrot1956}. 

The notion of fluctuating temperature has proved to be fundamental in the emerging field of quantum thermometry, where advances in nanotechnology now allow temperature sensing at sub-micron scales \cite{Giazotto2006a,Zanardi2008,Stace2010,DePasquale2015,Correa2015,Mehboudi2015,Jarzyna2015,Jevtic2015a,CarlosLuisDiasandPalacio2016,Johnson2016,Campbell2017,Hofer2017b,Puglisi2017}. 
Using the tools of quantum metrology \cite{Giovannetti2011a}, the relation~\eqref{eq:tur} can also be derived for weakly coupled quantum systems \cite{Zanardi2008,Stace2010,Correa2015}, where the equilibrium state is best described by the canonical ensemble.
Within the grand-canonical ensemble the impact of the indistinguishability of quantum particles on the estimation of temperature and the chemical potential has also been explored \cite{Marzolino2013}.
Relation~\eqref{eq:tur} informs us that when designing an accurate quantum thermometer one should search for systems with Hamiltonians that produce a large energy variance \cite{Correa2015}.

Recently there has been an emerging interest into the effects of strong coupling on temperature estimation \cite{Mehboudi2015,DePasquale2015,Correa2017}. Below the nanoscale the strength of interactions between the system and the reservoir may become non-negligible, and the local equilibrium state of the system will not be of Gibbs form \cite{Ludwig2010,Subasi2012}. In this regime thermodynamics needs to be adapted as the equilibrium properties of the system must now depend on the interaction energy \cite{Kirkwood1935,Ford1985,Jarzynski2004,Hanggi2008,Ingold2008,Gelin2009a,Hilt2011,Seifert2016,Jarzynski2016,Philbin2016,Strasberg2016a,Miller2017a,Strasberg2017}. We will see that the internal energy $U$ and its fluctuations $\lol U$ are determined by a modified internal energy operator, denoted by $\enmf$, that differs from the bare Hamiltonian of the system \cite{Seifert2016,Miller2017a}. This modification brings into question the validity of~\eqref{eq:tur} for general classical and quantum systems, and the aim of this paper is to investigate the impact of strong coupling on the thermodynamic uncertainty relation. 

Taking into account quantum properties of the effective internal energy operator and its temperature dependence, we here derive the general thermodynamic uncertainty principle valid at all coupling strengths. Formally this result follows from a general upper bound on the quantum Fisher information for exponential states. We prove that quantum fluctuations arising from coherences between energy states of the system lead to increased fluctuations in the underlying temperature. Most interestingly, the non-classical modifications to~\eqref{eq:tur} are quantified by the average Wigner-Yanase-Dyson skew information \cite{Wigner1963a,Luo2005,Li2011a,Frerot2016}, which is a quantity closely linked to measures of coherence, asymmetry and quantum speed limits \cite{Marvian2014a,PaivaPires2015}. We then demonstrate that the skew information is also linked to the heat capacity of the system through a modified fluctuation-dissipation relation. This result is used to find a new upper bound on the achievable signal-to-noise ratio of an unbiased temperature estimate, and we illustrate our bound with an example of a damped harmonic oscillator.

\section*{Results}


\subsection*{\bf The Wigner-Yanase-Dyson skew information}

Our analysis throughout the paper will rely on distinguishing between classical and non-classical fluctuations of observables in quantum mechanics, and we first present a framework for quantifying these different forms of statistical uncertainty for arbitrary mixed states.

Let us consider a quantum state $\hat{\rho}$ and an observable $\hat{A}$. Wigner and Yanase considered the problem of quantifying the quantum uncertainty in observable $\hat{A}$ for the case where $\hat{\rho}$ is mixed \cite{Wigner1963a}. However, they observed that the standard measure of uncertainty, namely the variance $\text{Var}[\hat{\rho},\hat{A}]:=\tr[\hat{\rho} \, \delta\hat{A}^2]$ with $\delta
\hat{A}=\hat{A}-\langle \hat{A} \rangle$, contains classical contributions due to mixing, and thus fails to fully quantify the non-classical fluctuations in the observable $\hat{A}$. This problem can be resolved by finding a quantum measure of uncertainty $Q[\hat{\rho},\hat{A}]$ and classical measure $K[\hat{\rho},\hat{A}]$ such that the variance can be partitioned according to 
\begin{align}\label{eq:var}
\text{Var}[\hat{\rho},\hat{A}]=Q[\hat{\rho},\hat{A}]+K[\hat{\rho},\hat{A}]. 
\end{align}
Following the framework introduced by Luo \cite{Luo2006}, these functions are required to fulfil three conditions: (i) both terms should be non-negative, $Q[\hat{\rho},\hat{A}]\geq 0$ and $K[\hat{\rho},\hat{A}]\geq 0$, so that they can be interpreted as forms of statistical uncertainty, (ii) if the state $\hat{\rho}$ is pure, then $Q[\hat{\rho},\hat{A}]=\text{Var}[\hat{\rho},\hat{A}]$ while $K[\hat{\rho},\hat{A}]=0$ as all uncertainty should be associated to quantum fluctuations alone, (iii) $Q[\hat{\rho},\hat{A}]$ must be convex with respect to $\hat{\rho}$, so that it decreases under classical mixing. Correspondingly, $K[\hat{\rho},\hat{A}]$ must be concave with respect to $\hat{\rho}$. 

The following function, known as the Wigner-Yanase-Dyson (WYD) skew information \cite{Wigner1963a} was shown to be a valid measure of quantum uncertainty:
\begin{align}\label{eq:WYD}
Q_a[\hat{\rho},\hat{A}]:=-\frac{1}{2}\tr\big[[\hat{A},\hat{\rho}^a][\hat{A},\hat{\rho}^{1-a}]\big]; \ \ \ \ a\in(0,1),
\end{align}
with the complementary classical uncertainty given by
\begin{align}
K_a[\hat{\rho},\hat{A}]:=\tr\big[\hat{\rho}^a \, \delta\hat{A} \ \hat{\rho}^{1-a} \delta\hat{A}\big]; \ \ \ \ a\in(0,1).
\end{align}
While conditions (i) and (ii) are easily verified, the convexity/concavity of $Q_a[\hat{\rho},\hat{A}]$ and $K_a[\hat{\rho},\hat{A}]$ respectively can be proven using Lieb's concavity theorem. 

The presence of the parameter $a$ demonstrates that there is no unique way of separating the quantum and classical contributions to the variance. We here follow the suggestion made in \cite{Li2011a,Frerot2016} and average over the interval $a\in(0,1)$ to define two new quantities:
\begin{align}\label{eq:qk}
&Q[\hat{\rho},\hat{A}]:=\int^1_0 da \ Q_a[\hat{\rho},\hat{A}], \\
&K[\hat{\rho},\hat{A}]:=\int^1_0 da \ K_a[\hat{\rho},\hat{A}].\label{eq:kk}
\end{align}

It is not only the $Q_a[\hat{\rho},\hat{A}]$ and $K_a[\hat{\rho},\hat{A}]$ that separate the quantum and classical fluctuations of a quantum observable $\hat{A}$ in a state $\hat{\rho}$ according to Eq.~\eqref{eq:var}, but also the averaged $Q[\hat{\rho},\hat{A}]$ and $K[\hat{\rho},\hat{A}]$. This follows from the linearity of the integrals in \eqref{eq:qk} and \eqref{eq:kk}  which also preserve  the conditions (i)-(iii). Throughout the remainder of the paper we will consider $Q[\hat{\rho},\hat{A}]$ and $K[\hat{\rho},\hat{A}]$ as the relevant measures of quantum and classical uncertainty, respectively. While this may appear to be an arbitrary choice, we will subsequently prove that the average skew information is intimately connected to thermodynamics.

\subsection*{\bf Bound on quantum Fisher information for exponential states}

We now prove that the average skew information is linked to the quality of a parameter estimate for a quantum exponential state. A quantum exponential state is of the form $\hatst=e^{-\hata}/Z_{\theta}$ where $Z_{\theta}=\tr[e^{-\hata}]$ and $\hata$ is a hermitian operator that is here assumed to depend analytically on a smooth parameter $\theta$. For any state of full rank, an operator $\hata$ can be found such that the state can be expressed in this form, i.e. all full rank states are exponential states.

We first recall the standard setup for estimating the parameter $\theta$ \cite{Caves1994}. First one performs a POVM measurement $\hat{M}(\xi)$, where $\int d\xi \ \hat{M}(\xi)=\hat{\mathbb{I}}$ and $\xi$ denotes the outcomes of the measurement which may be continuous or discrete. The probability of obtaining a particular outcome is $p(\xi |\theta)=\trace \hat{M}(\xi)\hatst]$. The measurement is repeated $n$ times with outcomes $\lbrace \xi_1, \xi_2, ..\xi_n \rbrace$, and one constructs a function $\tilde{\theta}=\tilde{\theta}(\xi_1, \xi_2, ..\xi_n)$ that estimates the true value of the parameter. We denote the average estimate by $\langle \tilde{\theta} \rangle$, where $\langle (..) \rangle=\int \ d\xi_1...d\xi_n \ p(\xi_1 |\theta)...p(\xi_n |\theta)(..)$, and assume the estimate is unbiased, ie. $\langle \tilde{\theta} \rangle=\theta$. In this case the mean-squared error in the estimate is equivalent to the variance, which is denoted by $\lol \theta^2=\langle \tilde{\theta}^2\rangle-\theta^2$.

The celebrated quantum Cram\'er-Rao inequality sets a lower bound on $\lol \theta$, optimised over all possible POVMs and estimator functions \cite{Holevo1982,Caves1994,Giovannetti2011a,Hayashi2017}:
\begin{align}\label{eq:cram}
\lol \theta \geq \frac{1}{\sqrt{nF(\theta)}},
\end{align}
where $F(\theta)$ is the quantum Fisher information (QFI). The bound becomes tight in the asymptotic limit $n\to\infty$ \cite{Giovannetti2011a}. If the exponential state belongs to the so-called `exponential family', which is true if $\hat{A}_\theta=\theta \hat{X}+\hat{Y}$ for commuting operators $\hat{X},\hat{Y}$, then the bound is also tight in the single-shot limit ($n=1$) \cite{Hayashi2017}.
The QFI with respect to $\theta$ is defined by $F(\theta):=\trace\hatst\hat{L}^2_\theta]$, where $\hat{L}_\theta$ is the symmetric logarithmic derivative which uniquely satisfies the operator equation $\partial_{ \theta} \hatst=\frac{1}{2}\lbrace \hat{L}_\theta, \hatst \rbrace$ \cite{Holevo1982}. Here $\lbrace \dots \rbrace$ denotes the anti-commutator.  

We now state a general upper bound on $F(\theta)$ valid for any exponential state:

\bigskip

\textit{Theorem 1:} For an exponential state $\hatst =e^{-\hata}/Z_{\theta}$  the QFI with respect to the parameter $\theta$ is bounded by
\begin{align}\label{thm:1}
	F(\theta)\leq K[\hatst,\hatb].
\end{align}
Here  $K[\hatst,\hatb]$ is defined in~\eqref{eq:kk}, and $\hatb$ is the hermitian observable $\hatb:=\partial_{ \theta}\hata$.
The bound becomes tight in the limits where $\hatst$ is maximally mixed. 
\smallskip

This theorem demonstrates that the strictly classical fluctuations in $\hatb$ constrain the achievable precision in estimates of $\theta$. The proof of the theorem is given in Appendix~\ref{app:a}.

We note that for states $\hatsigt$ that fulfil the von-Neumann equation $\partial_{\theta}\hatsigt = - i \, [A_{\theta}, \hatsigt]$ a connection between skew information $Q_{1/2}$ and parameter estimation has previously been made by Luo \cite{Luo2003}. While the particular dependence on $\theta$ implied by this equation is relevant for unitary parameter estimation \cite{Giovannetti2011a}, this dependence will not be relevant for temperature estimation since thermal states do not generally fulfil this von-Neumann equation. In contrast, we will see in the next section that Theorem~1 has implications for the achievable precision in determining temperature.

\subsection*{\bf Generalised thermodynamic uncertainty relation}
We will now use the results of the previous section to derive a new uncertainty relation between energy and temperature for a quantum system strongly interacting with a reservoir. To achieve this we will first discuss the appropriate energy operator for such a system, and then proceed to generalise~\eqref{eq:tur}.

A quantum system $\hilbs$ that interacts with a reservoir $\hilbr$ is described by a Hamiltonian 
\begin{align}\label{eq:hamtot}
\hamsr:=\hams\otimes\res{\hat{\mathbb{I}}}+\sys{\hat{\mathbb{I}}}\otimes\hamr+\vint,
\end{align}
where $\hams$ and $\hamr$ are the bare Hamiltonians of $\hilbs$ and $\hilbr$ respectively, while $\vint$ is an interaction term of arbitrary strength. We will consider situations where the environment is large compared to the system, i.e. the operator norms fulfil $||\hamr|| \gg ||\hams||, ||\vint||$. We make no further assumptions about the relative size of the coupling $||\vint||$ between the system and the environment, and the system's bare energy $||\hams||$. 
The global equilibrium state at temperature $T$ for the total Hamiltonian $\hilbsr$ is of Gibbs form $\statesr=\expo{\sr{\hat{H}}}/\partsr$ where $\beta=(k_{\text B} T)^{-1}$ and $\partsr=\tracesr\expo{\sr{\hat{H}}}]$ is the partition function for $\hilbsr$. The Boltzmann constant $k_{\text B}$ will be set to unity throughout. 

Due to the presence of the interaction term the reduced state of $\hilbs$, denoted $\states=\tracer \statesr]$, is generally not thermal with respect to $\hams$, unless the coupling is sufficiently weak, i.e. $||\hams|| \gg ||\vint||$. Therefore the partition function determined by $\hams$ can no longer be used to calculate the internal energy of the system \cite{Seifert2016}. To resolve this issue one can rewrite the state of $\hilbs$ as an effective Gibbs state $\states:=\expo{\hmf}/\parts$, where
\begin{align}\label{eq:hmf}
\hmf:=-\frac{1}{\beta}\lin\bigg(\frac{\tracer\expo{\sr{\hat{H}}}]}{\tracer \expo{\hamr}]}\bigg),
\end{align}
is the Hamiltonian of mean force \cite{Kirkwood1935,Jarzynski2004,Hanggi2008,Ingold2008,Gelin2009a,Hilt2011,Seifert2016,Jarzynski2016,Philbin2016,Miller2017a,Strasberg2017}. The operator $\hmf$ acts as a temperature-dependent effective Hamiltonian describing the equilibrium properties of $\hilbs$ through the effective partition function $\parts=\traces\expo{\hmf}]$. The free energy associated with $\parts$ also appears in the open system fluctuation relations \cite{Jarzynski2004,Campisi2009a}. 

The internal energy of $\hilbs$ can be computed from this partition function via $\sys{U}(T):=-\pbeta \ln \parts$. It is straightforward to show that $\sys{U}(T)$ is just the difference between the total energy, $\sr{U}=-\pbeta \ln \partsr$, and the energy of the reservoir, $\res{\tilde{U}}=-\pbeta \ln \partr$ with $\partr=\tracer\expo{\hamr}]$, in the absence of any coupling to $\hilbs$, i.e. $\sys{U}(T)=\sr{U}(T)-\res{\tilde{U}}(T)$. In other words, $\sys{U}(T)$ is the energy change induced from immersing the subsystem $\hilbs$ into the composite state $\hilbsr$ \cite{Ford1985,Jarzynski2016}.  

Seifert has remarked \cite{Seifert2016} that $\sys{U}(T)$ can be expressed as an expectation value,  $\sys{U}(T)=\average{\enmfT}$, of the following observable:
\begin{align}\label{eq:enmf}
\enmfT:=\deriv.
\end{align}
One can interpret $\enmfT$ as the effective energy operator describing the system, and we will refer to its eigenstates as ``the system energy states''. The introduction of this operator allows one to consider fluctuations in the energy $\lol \sys{U}=\sqrt{\text{Var}[\sys{\hat{\pi}},\enmf]}$. It is important to note that $\enmfT$ depends explicitly on the coupling $\vint$ and the temperature $T$. 

Our first observation is that, in general, $\enmfT$ differs from both the bare system Hamiltonian $\hams$ and the mean force Hamiltonian $\hmf$. Indeed, this effective energy operator for the system contains the bare energy part as well as an energetic contribution from the coupling, $ \enmfT = \hams + \partial_{\beta} [ \beta \, (\hmf - \hams)]$. 
Moreover, $\enmfT$ does not even commute with $\hams$ and $\hmf$. This non-commutativity implies that the state $\states$ exists in a superposition of energy states, aside from the trivial situation in which $[\hams+\hamr,\vint]=0$. As expected, in the limit of weak coupling $\enmfT$ reduces to the bare Hamiltonian $\hams$. 

\smallskip

We are now ready to state the generalised thermodynamic uncertainty relations for strongly coupled quantum systems. Following the approach taken by De Pasquale \textit{et al} \cite{DePasquale2015}, we consider the QFI $\sys{F}(\beta)$ associated with the inverse temperature $\beta$.  According to the quantum Cram\'er-Rao bound this functional quantifies the minimum extent to which the inverse temperature fluctuates from the perspective of $\hilbs$, and we denote these fluctuations by $\lol\sys{\beta}$. Given that the state of $\hilbs$ takes the form $\states:=\expo{\hmf}/\parts$ we can immediately apply Theorem 1 by identifying $\hatb=\enmfT$ with $\theta=\beta$, leading to $\sys{F}(\beta)\leq K[\sys{\hat{\pi}},\enmf]$. Applying~\eqref{eq:cram} for the single-shot case ($n=1$) and using the fact that $K[\sys{\hat{\pi}},\enmf]=\lol \sys{U}^2-Q[\sys{\hat{\pi}},\enmf]$, we obtain the following thermodynamic uncertainty relation:
\begin{align}\label{eq:result1}
\lol \sys{\beta} \geq \frac{1}{\sqrt{\lol \sys{U}^2-Q[\sys{\hat{\pi}},\enmf]}}  \geq \frac{1}{\lol \sys{U}}.
\end{align}
This is the main result of the paper and represents the strong-coupling generalisation of~\eqref{eq:tur}. 
It can be seen that the bound on the uncertainty in the inverse temperature is increased whenever quantum energy fluctuations are present. These additional fluctuations are quantified by the non-negative  $Q[\sys{\hat{\pi}},\enmf]$, increasing which implies a larger lower bound on $\lol \sys{\beta}$. One recovers the usual uncertainty relation when $Q[\sys{\hat{\pi}},\enmf]$ can be neglected, which is the case when the interaction commutes with the bare Hamiltonians of $\hilbs$ and $\hilbr$ or when the interaction is sufficiently weak. We note that $Q[\sys{\hat{\pi}},\enmf]$ vanishes for classical systems and~\eqref{eq:result1} reduces to the original uncertainty relation~\eqref{eq:tur}, but with energy fluctuations quantified by $\sys{\hat{E}}^*$ instead of the bare Hamiltonian $\hams$.

If one repeats the experiment $n$ times, then the uncertainty in the estimate can be improved by a factor of $1/\sqrt{n}$ \cite{Caves1994}. We remark that in the weak coupling limit, where $\hmf\simeq \hams$, the state of $\hilbs$ belongs to the exponential family, and hence the bound on $\lol \sys{\beta}$ becomes tight for a single measurement in agreement with~\eqref{eq:tur}. However, when $\vint$ is non-negligible the Hamiltonian of mean force cannot generally be expressed in the linear form $\beta\hmf=\beta \sys{\hat{X}}+\sys{\hat{Y}}$. This means in general it is necessary to take the asymptotic limit in order to saturate~\eqref{eq:result1}.

\subsection*{\bf Fluctuation-Dissipation relation beyond weak-coupling}

We now detail the impact of strong interactions on the heat capacity of the quantum system and the implications for the precision of temperature measurements. For a fixed volume of the system, the heat capacity is defined as the temperature derivative of the internal energy $\sys{U}(T)$ \cite{Hanggi2008,Ingold2008}, i.e.
\begin{align}
\sys{C}(T):=\frac{\partial \sys{U}}{\partial T}.
\end{align}
In standard thermodynamics where the system is described by a Gibbs state the fluctuation-dissipation relation (FDR) states that the heat capacity is proportional to the fluctuations in energy, i.e. $\sys{C}(T)=\lol \sys{U}^2/T^2$. However, example studies of open quantum systems of the form~\eqref{eq:hamtot} have shown that the heat capacity can become negative at low temperatures \cite{Hanggi2008,Ingold2008,Campisi2009b,Campisi2010b}, thus implying it cannot be proportional to a positive variance in general. 

Our second result indeed shows that there are two additional contributions to the fluctuation-dissipation relation due to strong-coupling (see Appendix~\ref{app:b}):
\begin{align}\label{eq:heatcap}
\sys{C}(T)=\frac{\lol \sys{U}^2}{T^2}-\frac{Q[\sys{\hat{\pi}},\enmf]}{T^2}+\big\langle\partial_{\scriptscriptstyle T}\enmf\big\rangle,
\end{align}
implying that $\sys{C}(T)$ can be less than $\lol \sys{U}^2/T^2$ and even negative. We see that the first correction is due to the quantum fluctuations in energy given by the average WYD information $Q[\sys{\hat{\pi}},\enmf]$, which only vanishes in the classical limit where $[\enmfT,\states]=0$. The second correction is a dissipation term stemming from the temperature dependence of the internal energy operator~\eqref{eq:enmf}. Notably this term can still be present in the classical limit where the energy operator may depend on temperature if the coupling is non-negligible. As expected both terms can be dropped in the limit of vanishing coupling and the standard FDR is recovered.  

\subsection*{\bf Bound on signal-to-noise ratio for  temperature estimates}

Let us denote the uncertainty in the temperature from a given unbiased estimation scheme by $\lol \sys{T}$, with measurements performed on $\hilbs$ alone. It is known \cite{Zanardi2008,Stace2010,Jahnke2011,Correa2015} that in the weak-coupling limit, the optimal signal-to-noise ratio for estimating $T$ from a single measurement is bounded by $\sys{C}(T)$:
\begin{align}\label{eq:tur2}
\bigg(\frac{T}{\lol \sys{T}}\bigg)^2\leq \sys{C}(T).
\end{align}
This bound is tight for a single measurement of $T$ and implies that precise measurements of the temperature require a large heat capacity. The result follows straightforwardly from the quantum Cram\'er-Rao inequality and the standard FDR.

Using our modified FDR~\eqref{eq:heatcap}, we here give the strong-coupling generalisation of the bound~\eqref{eq:tur2}. Considering estimates of $T$ rather than the inverse temperature $\beta$, a simple change of variables reveals that the QFI with respect to $T$ is related to that of $\beta$, $\sys{F}(\beta)=T^4 \sys{F}(T)$. From Theorem 1 we again have $T^4 \sys{F}(T)\leq K[\sys{\hat{\pi}},\enmf]$, and combining this with~\eqref{eq:heatcap} and~\eqref{eq:cram} we obtain:
\begin{align}\label{eq:result2}
\bigg(\frac{T}{\lol \sys{T}}\bigg)^2\leq \sys{C}(T)-\big\langle\partial_{\scriptscriptstyle T}\enmf\big\rangle.
\end{align}
This is our third result and demonstrates that the optimal signal-to-noise ratio for estimating the temperature of $\hilbs$ is bounded by both the heat capacity and the added dissipation term, which can be both positive or negative. This bound is independently tight in both the high temperature and weak-coupling limits. In these regimes the POVM saturating~\eqref{eq:result2} is given by the maximum-likelihood estimator measured in the basis of the relevant symmetric logarithmic derivative \cite{Caves1994}. We stress that~\eqref{eq:result2} is valid in the classical limit, in which case it is always tight. We remark that the RHS of~\eqref{eq:result2} can alternatively be expressed in terms of the skew information, in which case $(T/\lol \sys{T})^2\leq \lol \sys{U}^2/T^2-Q[\sys{\hat{\pi}},\enmf]/T^2$.

\subsection*{Application to damped harmonic oscillator} 

\begin{figure}[t]
\includegraphics[trim=0cm 0cm 0cm 0cm,clip, width=0.45\textwidth]{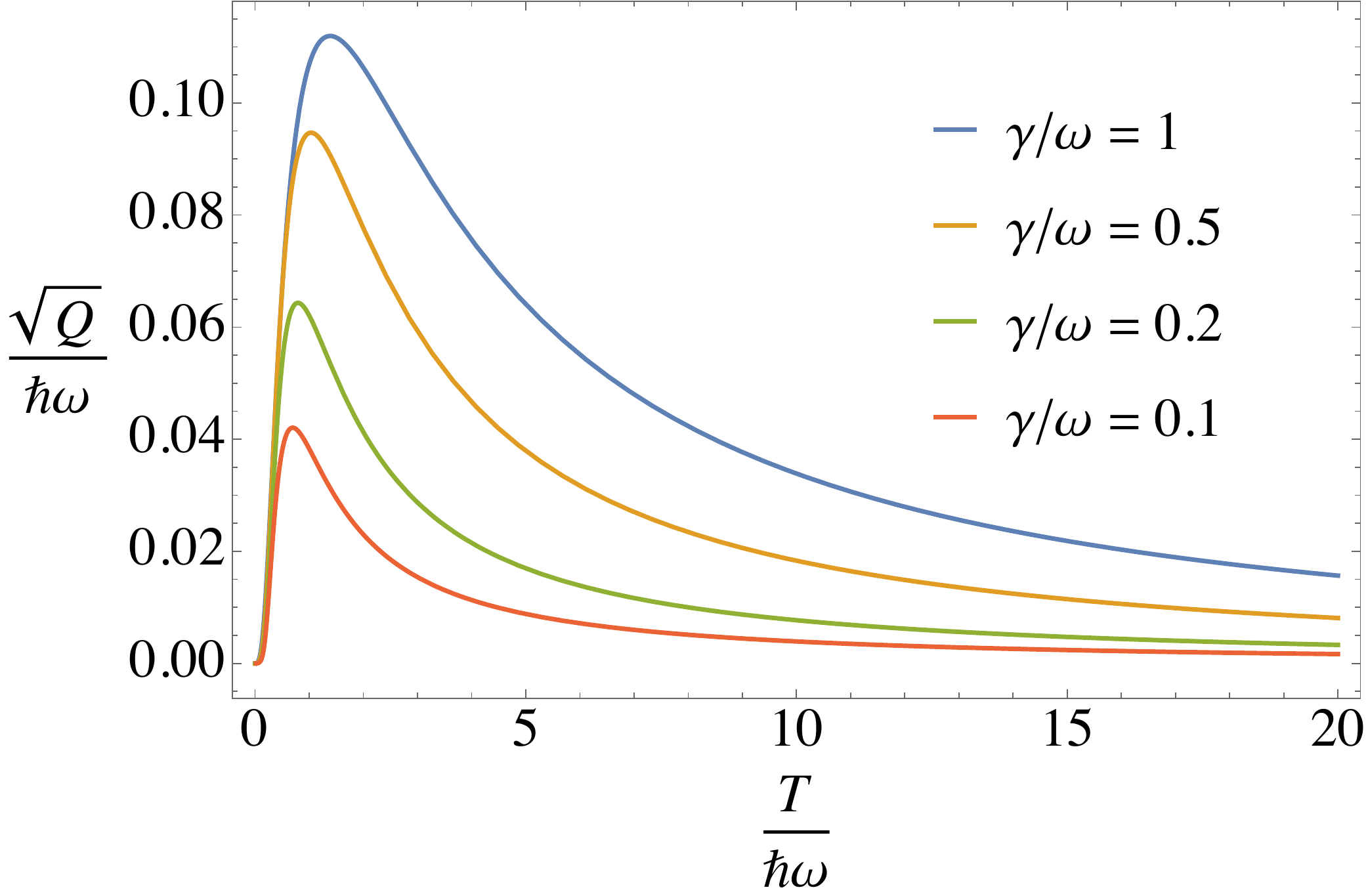}
\caption{{\bf Skew information for the damped oscillator.} Plot of quantum energetic fluctuations $\sqrt{Q[\sys{\hat{\pi}},\enmf]}/\hbar\omega$ for the damped oscillator as a function of $T/\hbar\omega$ for different coupling strengths $\gamma$. Here $Q[\sys{\hat{\pi}},\enmf]$ is the average Wigner-Yanase-Dyson skew information for the effective energy operator $\enmf$. These fluctuations are present when the state of the oscillator $\states$ is not diagonal in the basis of $\enmf$ due to the non-negligible interaction between the system and reservoir. The plot shows that increasing the coupling $\gamma$ leads to an increase in the skew information. The quantum fluctuations are most pronounced at low temperatures where the thermal energies become comparable to the oscillator spacing, $T\simeq \hbar\omega$. As expected, the skew information decreases to zero in both the high temperature and weak coupling limits.}
\label{fig:skew}
\end{figure}

\begin{figure}[t]
\centering
\includegraphics[trim=5.5cm 2cm 1.8cm 3.6cm,clip,width=0.50\textwidth]{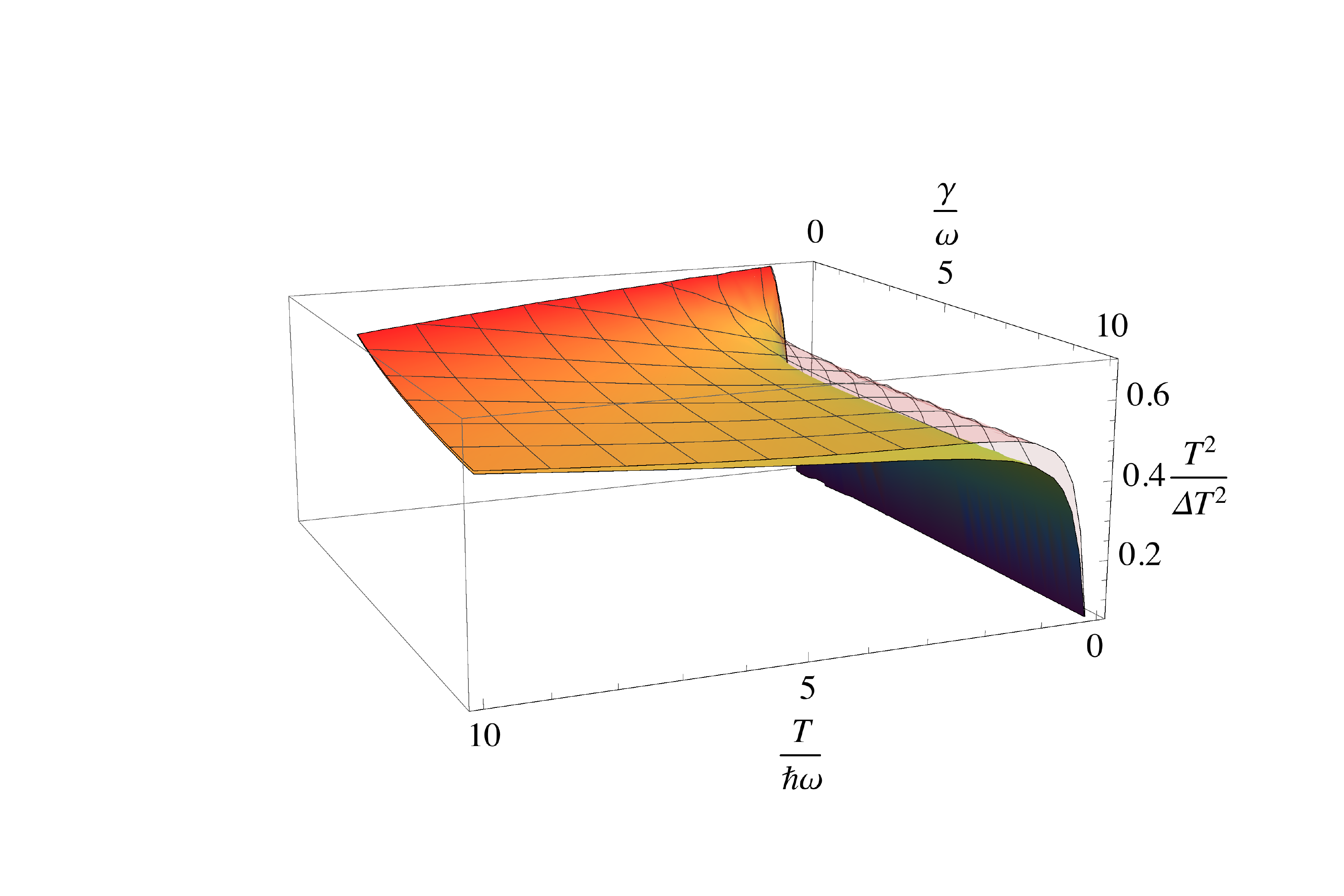}
\caption{{\bf Bound on temperature signal-to-noise ratio.} The coloured plot shows the optimal signal-to-noise ratio $(T/\lol \sys{T})_{\scriptscriptstyle \text{opt}}^2$ of an unbiased temperature estimate for the damped oscillator, as a function of temperature $T$ and coupling strength $\gamma$. This optimal measurement is determined by the quantum Fisher information, which places an asymptotically achievable lower bound on the temperature fluctuations $\lol \sys{T}$ through the Cram\'er-Rao inequality \cite{DePasquale2015}. The mesh plot shows the upper bound on $(T/\lol \sys{T})_{\scriptscriptstyle \text{opt}}^2$ derived here from the generalised thermodynamic uncertainty relation~\eqref{eq:result2}. This uncertainty relation links the temperature fluctuations to the heat capacity of the system at arbitrary coupling strengths. It can be seen that the upper bound becomes tight in both the high temperature and weak coupling limits.}
\label{fig:QFI}
\end{figure}

While the bound~\eqref{eq:result2} is tight in the high temperature limit, for general open quantum systems the accuracy of the bound is not known. We show that the bound is very good for the example of a damped harmonic oscillator linearly coupled to $N$ harmonic oscillators in the reservoir \cite{Grabert1984,Hilt2011,Pucci2013}. Experimentally, such a model  describes the behaviour of nano-mechanical resonators \cite{Connell2010} and BEC impurities \cite{Lampo2017}. Here the system Hamiltonian is $\hams=\frac{\hp^2}{2M}+\frac{M\omega^2 \hx^2}{2}$, while the reservoir Hamiltonian is $\hamr=\sum^N_{j=1}\big(\frac{\hp_j^2}{2M_j}+\frac{M_j\omega_j^2 \hx^2_j}{2}\big)$ and the interaction term is given by
\begin{align}
\vint=\sum^N_{j=1}\bigg(-\lambda_j\hx\otimes\hx_j+\frac{\lambda^2_j}{2M_j\omega^2_j}\hx^2\bigg).
\end{align}
To allow a fully analytical solution, the reservoir frequencies are chosen equidistant, $\omega_j=j\Delta$ and the continuum limit is taken so that $\Delta\rightarrow 0$ (and $N\rightarrow \infty$). The coupling constants are chosen as the Drude-Ullersma spectrum \cite{Hilt2011},
$\lambda_j=\sqrt{\frac{2\gamma M_j M\omega_j^2 \Delta}{\pi}\frac{\omega_{\text{D}}^2}{\omega_{\text{D}}^2+\omega_j^2}}$, where $\gamma$ is the damping coefficient controlling the interaction strength and $\omega_{\text{D}}$ is a large cutoff frequency.

As shown by Grabert {\it et al.} \cite{Grabert1984}, the resulting Hamiltonian of mean force for the oscillator can be parameterised by a temperature-dependent mass and frequency,
\begin{align}
	\hmf=\frac{\hp^2}{2\mt}+\frac{\mt\omt^2 \hx^2}{2} = \hbar \omt \, \left(\hat{n}_{\scriptscriptstyle T}+\frac{1}{2}\right),
\end{align}
where $\mt$ and $\omt$ are given through the expectation values of $\hp^2$ and $\hx^2$ in the global thermal state, see Appendix~\ref{app:c} for detailed expressions. In its diagonal form the mean-force Hamiltonian contains a temperature-dependent number operator, $\hat{n}_{\scriptscriptstyle T}=\ha^{\dagger}\ha$, with annihilation operator  $\ha=\sqrt{\frac{\hA}{2\hbar}}(\hx+\frac{i}{\hA} \, \hp)$ with $\hA =\mt \, \omt$.

The internal energy operator is now obtained by straightforward differentiation, see \eqref{eq:enmf}, and given by
\begin{align}
	\enmfT=\hD \hmf-\hG\frac{\ha^2+(\ha^{\dagger})^2}{2}, 
\end{align}
where $\hD=1-\frac{\omt'}{\omt}T$ and $\hG=\hbar\omt T\frac{\hA'}{\hA}$. Using this operator we obtain analytic expressions for $\sys{C}(T)$, $\sys{F}(T)$, $Q[\sys{\hat{\pi}},\enmf]$ and $\langle\partial_{\scriptscriptstyle T} \enmf\rangle$ in Appendix~\ref{app:c}. 


Figure~\ref{fig:skew} shows the square root of the average skew information $Q[\sys{\hat{\pi}},\enmf]$ in units of $\hbar\omega$ as a function of temperature for different coupling strengths. As expected we see that the quantum fluctuations in energy vanish in the high temperature limit, while fluctuations grow with increased coupling strengths due to increased non-commutativity between $\enmfT$ and the state $\sys{\hat{\pi}}$ of the oscillator. Interestingly we see that $Q[\sys{\hat{\pi}},\enmf]$ decays exponentially to zero in the low temperature limit, implying that the state of the oscillator commutes with the internal energy operator in this regime. Whether this is a general feature or specific to the example here remains an open question.

Figure~\ref{fig:QFI} shows the optimal signal-to-noise ratio for estimating $T$  determined by the Cram\'er-Rao bound~\eqref{eq:cram}, $(T/\lol \sys{T})^2_{\scriptscriptstyle \text{opt}}=T^2\sys{F}(T)$, as a function of temperature $T$ and coupling strength $\gamma$. The bound we derived in~\eqref{eq:result2} given by the heat capacity and an additional dissipation term is also plotted and shows very good agreement with the optimum estimation scheme quantified by the quantum Fisher information. The bound clearly becomes tight in the high-temperature limit ($T\rightarrow\infty$) independent of the coupling strength. Conversely the bound is also tight in the weak-coupling limit ($\gamma\rightarrow 0$) independent of the temperature. The optimum and the bound both converge exponentially to zero as $T\rightarrow 0$, albeit with different rates of decay. Outside of these limits the difference between the bound and $(T/\lol \sys{T})^2_{\scriptscriptstyle \text{opt}}$ has a maximum, and at the temperature and coupling for which this maximum occurs the bound is roughly $30\%$ greater than $(T/\lol \sys{T})^2_{\scriptscriptstyle \text{opt}}$.

\section*{Discussion}

In this paper we have shown how non-negligible interactions influence fluctuations in temperature at the nanoscale. Our main result~\eqref{eq:result1} is a thermodynamic uncertainty relation extending the well-known complementarity relation~\eqref{eq:tur} between energy and temperature to all interaction strengths. This derivation is based on a new bound on the quantum Fisher information for exponential states which we prove in Theorem 1. As Theorem 1 is valid for any state of full-rank, the bound will be of interest to other areas of quantum metrology. Our uncertainty relation shows that for a given finite spread in energy, unbiased estimates of the underlying temperature are limited to a greater extent due to coherences between energy states. These coherences only arise for quantum systems beyond the weak coupling assumption. We found that these additional temperature fluctuations are quantified by the average Wigner-Yanase-Dyson skew information, thereby establishing a new link between quantum and classical forms of statistical uncertainty in nanoscale thermodynamics. With coherence now understood to be an important resource in the performance of small-scale heat engines \cite{Uzdin2015,Kammerlander2016}, our findings suggest that the skew information could be used to unveil further non-classical aspects of quantum thermodynamics. This complements previous results that connect skew information to both unitary phase estimation \cite{Luo2003} and quantum speed limits \cite{PaivaPires2015}.

Our second result~\eqref{eq:heatcap} is a generalisation of the well-known fluctuation-dissipation relation to systems beyond the weak coupling regime. This further establishes a connection between the skew information and the system's heat capacity $\sys{C}(T)$. Proving that the heat capacity, with its strong coupling corrections, vanishes in the zero-temperature limit in accordance with the third law of thermodynamics remains an open question. The appearance of the skew information in~\eqref{eq:heatcap} suggests that quantum coherences may play a role in ensuring its validity.
 Recent resource-theoretic derivations of the third law \cite{Masanes2017,Wilming2017b} could provide a possible avenue for exploring the impact of coherences. 

By applying the fluctuation-dissipation relation to temperature estimation we derive our third result, an upper bound~\eqref{eq:result2} on the optimal signal-to-noise ratio expressed in terms of the system's heat capacity. Notably the bound implies that when designing a probe to measure $T$, its bare Hamiltonian and interaction with the sample should be chosen so as to both maximise $\sys{C}(T)$ whilst minimising the additional dissipation term $\big\langle\partial_{\scriptscriptstyle T}\enmf\big\rangle$. It is an interesting open question to consider the form of Hamiltonians that achieve this optimisation in the strong coupling scenario. Furthermore, one expects that improvements to low-temperature thermometry resulting from strong interactions, such as those observed in \cite{Correa2017}, will be connected to the properties of the effective internal energy operator. In particular, it is clear from~\eqref{eq:result2} that any improved scaling of the QFI at low temperatures must be determined by the relative scaling of $\sys{C}(T)$ and $\big\langle\partial_{\scriptscriptstyle T}\enmf\big\rangle$, and exploring this further remains a promising direction of research. Advancements in nanotechnology now enable temperature sensing over microscopic spatial resolutions \cite{Neumann2013,Kucsko2013}, and understanding how to exploit interactions between a probe and its surroundings will be crucial to the development of these nanoscale thermometers.

The presented approach opens up opportunities for exploring the intermediate regime between the limiting cases \cite{Gonzalez2017,Hofer2017c} of standard thermodynamics with negligible interactions and those where correlations play a prominent role \cite{Iles-Smith2014,Strasberg2016a,Newman2017}. The results establish a new connection between abstract measures of quantum information theory, such as the quantum Fisher information and skew information, and a material's effective thermodynamic properties. This provides a starting point for future investigations into nanoscale thermodynamics, extending into the regime where the weak coupling assumption is not justified.

\bibliographystyle{apsrev4-1}
\bibliography{mainbib.bib}

\widetext

\appendix

\section{Proof of~\ref{thm:1}}\label{app:a}

We begin by considering an exponential state $\hatst$ dependent on smooth parameter $\theta$ of the form $\hat{\rho}(\theta)=e^{-\hata}/Z$, where $Z=\tr[e^{-\hata}]$ and $\hata$ is some positive hermitian operator. Suppressing the dependence on $\theta$ for now, let us denote the spectral decomposition by $\hatst=\sum_n p_n \ket{\psi_n}\bra{\psi_n}$ where the eigenstates satisfy $\hata\ket{\psi_n}=\lambda_n\ket{\psi_n}$. We arrange the sum in decreasing order, so that $p_n\geq p_m$ if $n<m$. The QFI with respect to $\theta$ is then $F(\theta):=\tr[\hatst \hat{L}^2_\theta]$, where $\hat{L}_\theta$ is the operator satisfying 
\begin{align}\label{eq:logdiv}
\partial_\theta \hatst=\frac{1}{2}\lbrace \hatst, \hat{L}_\theta \rbrace.
\end{align}
By expanding both sides of~\eqref{eq:logdiv} in the basis of $\hatst$, one may show that the QFI can be written as follows \cite{Liu2014a}:
\begin{align}\label{eq:fishexp}
F(\theta)=2\sum_{n,m}\frac{|\bra{\psi_n} \partial_\theta \hatst\ket{\psi_m}|^2}{p_n+p_m}.
\end{align}
We define the operator $\hatb=\partial_\theta \hata$ and note that $\text{Var}[\hatst,\hatb]=\tr[\delta \hatb^2 \hatst]$ where $\delta \hatb=\hatb+\partial_\theta \ln Z$. Given the exponential form of $\hatst$, we use the following integral expression to expand the derivative \cite{Wilcox1967a}:
\begin{align}\label{eq:int}
\partial_\theta \ e^{-\hat{A}(\theta)}:=-\int^1_0 da \ e^{-(1-a) \hat{A}(\theta)}\partial_\theta [\hat{A}(\theta)]e^{-a \hat{A}(\theta)},
\end{align}
where $a\in\mathbb{R}$ is a real number. Using this the QFI becomes
\begin{align}\label{eq:part1}
\nonumber F(\theta)&=2\sum_{n,m}\frac{|\bra{\psi_n} \partial_\theta e^{-(\hata+\ln Z)}\ket{\psi_m}|^2}{p_n+p_m}, \\
\nonumber&=\frac{2}{Z^2}\sum_{n,m}\frac{1}{p_n+p_m}\bigg|\bra{\psi_n}\int^1_0 da \ e^{-(1-a)\hata}\delta \hatb  \ e^{-a\hata} \ket{\psi_m}\bigg|^2, \\
\nonumber&=\sum_n p_n\big|\bra{\psi_n}\delta\hatb \ket{\psi_n}\big|^2+\frac{4}{Z^2}\sum_{n< m}\frac{1}{p_n+p_m}\big|\bra{\psi_m}\hatb\ket{\psi_n}\big|^2 \bigg[\int^1_0 da \ e^{-\big(a\lambda_m+(1-a)\lambda_n\big)}\bigg]^2, \\
&=\sum_n p_n\big|\bra{\psi_n}\delta\hatb\ket{\psi_n}\big|^2+4\sum_{n< m} \frac{(p_n-p_m)^2}{p_n+p_m}\big|\bra{\psi_n}\hatb\ket{\psi_m}\big|^2\frac{1}{(\ln p_n-\ln p_m)^2}, 
\end{align}
Let us now use the following expression for the variance:
\begin{align}\label{eq:var2}
\nonumber\text{Var}[\hatst,\hatb]&=\sum_{n,m}\frac{p_n+p_m}{2} \big|\bra{\psi_n}\delta\hatb\ket{\psi_m}\big|^2, \\
\nonumber&=\sum_{n}p_n\big|\bra{\psi_n}\delta\hatb\ket{\psi_n}\big|^2+ \sum_{n\neq m}\frac{p_n+p_m}{2} \big|\bra{\psi_n}\delta\hatb\ket{\psi_m}\big|^2, \\
&=\sum_{n}p_n\big|\bra{\psi_n}\delta\hatb\ket{\psi_n}\big|^2+ \sum_{n<m}(p_n+p_m)\big|\bra{\psi_n}\delta\hatb\ket{\psi_m}\big|^2, 
\end{align}
Comparing this with~\eqref{eq:part1} we now add and subtract the sum $\sum_{n<m}(p_n+p_m)\big|\bra{\psi_n}\delta\hatb\ket{\psi_m}\big|^2$ to the RHS of~\eqref{eq:part1}, obtaining
\begin{align}\label{eq:fish2}
F(\theta)=\text{Var}[\hatst,\hatb]+\sum_{n< m} \bigg[\bigg(\frac{2(p_n-p_m)}{\ln (p_n/p_m)}\bigg)\bigg(\frac{2(p_n-p_m)}{(p_n+p_m)\ln (p_n/p_m)}\bigg)-(p_n+p_m)\bigg]\big|\bra{\psi_n}\hatb\ket{\psi_m}\big|^2,
\end{align}
We now turn to the average WYD skew information of observable $\hatb$, which is given by~\eqref{eq:qk}:
\begin{align}
Q[\hatst,\hatb]=-\frac{1}{2} \int^1_0 da \ \tr\big[[\hatb,\hatst^a][\hatb,\hatst^{1-a}]\big].
\end{align}
It follows from the analysis of \cite{Li2011a} that for a full-rank state $Q[\hatst,\hatb]$ can also be expanded in the eigenbasis of $\hatst$, yielding
\begin{align}\label{eq:part2}
Q[\hatst,\hatb]=\sum_{n<m}\bigg(p_n+p_m-\frac{2(p_n-p_m)}{\ln p_n -\ln p_m}\bigg)\big|\bra{\psi_n}\hatb\ket{\psi_m}\big|^2.
\end{align}
We now bound the QFI following on from~\eqref{eq:fish2}:
\begin{align}\label{eq:part3}
\nonumber F(\theta)&=\text{Var}[\hatst,\hatb]+\sum_{n< m} \bigg[\bigg(\frac{2(p_n-p_m)}{\ln (p_n/p_m)}\bigg)\bigg(\frac{2(p_n-p_m)}{(p_n+p_m)\ln (p_n/p_m)}\bigg)-(p_n+p_m)\bigg]\big|\bra{\psi_n}\hatb\ket{\psi_m}\big|^2, \\
\nonumber&\leq\text{Var}[\hatst,\hatb]+\sum_{n< m} \bigg[\frac{2(p_n-p_m)}{\ln (p_n/p_m)}-(p_n+p_m)\bigg]\big|\bra{\psi_n}\hatb\ket{\psi_m}\big|^2, \\
\nonumber&= \text{Var}[\hatst,\hatb]-Q[\hatst,\hatb], \\
&=K[\hatst,\hatb],
\end{align}
where in the second line we used the fact that $(p_n-p_m)/\ln (p_n/p_m)\geq 0$ since $p_n \geq p_m$ for $n<m$, and the inequality 
\begin{align}
\frac{x-1}{x+1}\leq \ln\sqrt{x}; \ \ \ \ x\geq 1,
\end{align}
identifying $x=p_n/p_m\geq 1$. This allowed us to use
\begin{align}
\bigg(\frac{2(p_n-p_m)}{(p_n+p_m)\ln (p_n/p_m)}\bigg)\leq 1, 
\end{align}
for each term inside the sum. In the third line we used the expression~\eqref{eq:part2} for the skew information, and we employed~\eqref{eq:var} in the final line. This completes the proof of Theorem~\ref{thm:1}. 

\section{Derivation of~\eqref{eq:heatcap}}\label{app:b}

Denote the operator $\delta\enmf:=\enmfT-\average{\enmfT}$ as the deviation in internal energy, dropping the temperature dependence for now. Using~\eqref{eq:qk} we now evaluate the average WYD skew information of the internal energy:
\begin{align}
\nonumber Q[\sys{\hat{\pi}},\enmf]&=\text{Var}[\sys{\hat{\pi}},\enmf]-K[\sys{\hat{\pi}},\enmf], \\
\nonumber&=\text{Var}[\sys{\hat{\pi}},\enmf]-\int^1_0 da \ \trace\sys{\hat{\pi}}^{1-a}\delta\enmf \ \sys{\hat{\pi}}^a \ \delta\enmf], \\
\nonumber&=\text{Var}[\sys{\hat{\pi}},\enmf]-\int^1_0 da \ \trace e^{-(1-a)(\beta\sys{\hat{H}}^*+\ln \sys{Z}^*)}\delta\enmf e^{-a(\beta\sys{\hat{H}}^*+\ln \sys{Z}^*)}\delta\enmf], \\
\nonumber&=\text{Var}[\sys{\hat{\pi}},\enmf]+\trace \delta\enmf \ \partial_\beta \sys{\hat{\pi}}   ], \\
\nonumber&=\text{Var}[\sys{\hat{\pi}},\enmf]+\trace \enmf \ \partial_\beta \sys{\hat{\pi}} ], \\
\nonumber&=\text{Var}[\sys{\hat{\pi}},\enmf]-T^2\trace \enmf \ \partial_{\scriptscriptstyle T} \sys{\hat{\pi}} ], \\
\nonumber&=\text{Var}[\sys{\hat{\pi}},\enmf]-T^2\partial_{\scriptscriptstyle T}  \trace \enmf \ \sys{\hat{\pi}} ]+T^2\trace \partial_{\scriptscriptstyle T}\enmf \ \sys{\hat{\pi}} ], \\
&=\text{Var}[\sys{\hat{\pi}},\enmf]- T^2 \sys{C}(T)+T^2\big\langle\partial_{T}\enmf\big\rangle,
\end{align}
where we used the relation $\delta \enmf=\partial_{\beta}(\beta\sys{\hat{H}}^*+\ln \sys{Z}^*)$ and~\eqref{eq:int} in the the fourth line, and the fact that the operator $\partial_\beta \sys{\hat{\pi}}$ is traceless in the fifth line. Rearranging both sides completes the derivation of~\eqref{eq:heatcap}.

\section{Example}\label{app:c}

As stated in the main text the mean force Hamiltonian of the probe is given by $\hmf=\omt(\hat{n}_{\scriptscriptstyle T}+\frac{1}{2})$, with $\hat{n}_{\scriptscriptstyle T}=\ha^{\dagger}\ha$ and $\ha=\sqrt{\frac{\hA}{2}}(\hx+i\frac{\hp}{\hA})$. We will set $\hbar=1$ throughout and $\hA$ is given by $\hA=\sqrt{\langle\hp^2 \rangle/\langle\hx^2\rangle}$, where the effective mass and frequency are given respectively by 
\begin{align}
&\mt=\omt^{-1}\sqrt{\frac{\langle\hp^2 \rangle}{\langle\hx^2\rangle}}, \\
&\omt=2T \ \text{arcoth}(2\sqrt{\langle\hp^2\rangle\langle\hx^2\rangle}).
\end{align}
We can diagonalise the state of the probe in terms of the number states of $\hat{n}_{\scriptscriptstyle T}$, so $\states=e^{-\beta \hmf}/\sys{Z}^*=\sum^{\infty}_{n=0}p_n \ket{n}\bra{n}$ where
\begin{align}
p_n=\frac{e^{-\beta \epsilon_n}}{\sys{Z}^*}, \ \ \ \ \sys{Z}^*=2 \ \text{sinh}^{-1}\bigg (\bom\bigg ),
\end{align}
and $\epsilon_n=\omt (n+\frac{1}{2})$. Furthermore, from the main text the internal energy operator is given by 
\begin{align}
\enmfT=\hD \hmf-\hG\frac{\ha^2+(\ha^{\dagger})^2}{2}, 
\end{align}
with $\hD=1-\frac{\omt'}{\omt}T$ and $\hG=\omt T\frac{\hA'}{\hA}$. The functions $\hG$ and $\hD$ are determined by the effective mass and frequency of the oscillator defined above, so we need to calculate $\langle\hx^2\rangle$ and $\langle \hp^2\rangle$. In the continuum limit $N\rightarrow \infty$ the exact expressions for the quadratures are found to be \cite{Grabert1984}:
\begin{align}
&\langle\hx^2\rangle= \frac{1}{M\beta\omega^2}+\frac{\hbar}{M\pi} \sum_{i=1}^3 \bigg[ \frac{(\lambda_i-\omega_D) \Gamma(1+\frac{\beta\hbar\lambda_i}{2\pi})}{(\lambda_{i+1}-\lambda_i)(\lambda_{i-1}-\lambda_i)}\bigg],\\
&\langle \hp^2\rangle=M\omega^2\langle\hx^2\rangle+\frac{\hbar M\gamma\omega_D}{\pi}\sum_{i=1}^3 \bigg[ \frac{\lambda_i\Gamma(1+\frac{\beta\hbar\lambda_i}{2\pi})}{(\lambda_{i+1}-\lambda_i)(\lambda_{i-1}-\lambda_i)}\bigg], 
\end{align}
where $\Gamma(z)$ is the digamma function and $\lambda_i$ are the characteristic frequencies of the oscillator. In the limit of a large cutoff frequency, $\omega_D \gg \omega,\gamma$ the frequencies are given by
\begin{align}
\nonumber &\lambda_1= \frac{\gamma}{2}+\sqrt{\frac{\gamma^2}{4}-\omega^2}, \\
\nonumber &\lambda_2=  \frac{\gamma}{2}- \sqrt{\frac{\gamma^2}{4}-\omega^2}, \\
&\lambda_3= \omega_D-\gamma.
\end{align}
Due to the complicated dependence on $T$ we will not present the exact analytic expressions for $\hG$ and $\hD$, but we will proceed to calculate $\sys{C}(T)$, $\sys{F}(T)$ and $\langle\partial_{\scriptscriptstyle T} \enmf\rangle$. The average internal energy is found to be
\begin{align}
&\langle \enmf \rangle=\hD\langle \hmf \rangle, \\
&\langle \hmf \rangle= \frac{\omt}{2}\text{coth}\bigg(\bom\bigg).
\end{align}
The heat capacity can now be calculated by differentiating the average energy:
\begin{align}
\nonumber \sys{C}(T)&=\partial_{\scriptscriptstyle T}\langle \enmf \rangle, \\
\nonumber &=\hD' \frac{\omt}{2}\text{coth}\bigg(\bom\bigg)+\hD \frac{\omt'}{2}\text{coth}\bigg(\bom\bigg)-\frac{\hD\omt(\omt'\beta-\omt\beta^2)}{4\text{sinh}^{2}\bigg (\bom\bigg )}, \\
&=\frac{1}{2}\text{coth}\bigg(\bom\bigg)(\hD'\omt+\hD\omt')-\frac{\hD\omt\beta(\omt'-\omt\beta)}{4\text{sinh}^{2}\bigg (\bom\bigg )}.
\end{align}
In order to calculate the QFI~\eqref{eq:fishexp} and skew information~\eqref{eq:part2} we will need to obtain the elements $E_{nm}=|\bra{n}\delta\hat{E}^*\ket{m}|^2$, where $\delta\hat{E}^*=\enmf-\langle\enmf \rangle$. Firstly one finds the following:
\begin{align}
\nonumber \bra{n}\delta\hat{E}^*\ket{m}&=\bra{n}\enmf\ket{m}-\langle \enmf \rangle\delta_{n,m}, \\
\nonumber &=\hD\bra{n}\hmf\ket{m}-\hG\bra{n}\frac{\ha^2+(\ha^{\dagger})^2}{2}\ket{m}-\langle \enmf \rangle\delta_{n,m}, \\
&=\hD(\epsilon_n-\langle\hmf \rangle)\delta_{n,m}-\hG\bigg(\frac{\sqrt{m}\sqrt{m-1}}{2}\delta_{n,m-2}+\frac{\sqrt{n}\sqrt{n-1}}{2}\delta_{m,n-2}\bigg),
\end{align}
where $\delta_{n,m}$ represents the Kronecker-Delta function. Squaring both sides yields
\begin{align}
E_{nm}=\hD^2(\epsilon_n-\langle\hmf\rangle)^2\delta_{n,m}+\frac{\hG^2}{4}\bigg((n+2)(n+1)\delta_{n+2,m}+(m+2)(m+1)\delta_{m+2,n}\bigg),
\end{align}
The variance in internal energy is given by~\eqref{eq:var2}, so that
\begin{align}
\nonumber  \text{Var}[\sys{\hat{\pi}},\enmf]&=\sum_{n,m=0}^{\infty}p_n E_{nm}, \\
\nonumber &=\hD^2\text{Var}[\sys{\hat{\pi}},\sys{\hat{H}}^*]+\frac{\hG^2}{2}\sum_{n=0}^{\infty}p_n(n+1)(n+2), \\
\nonumber &=\hD^2\text{Var}[\sys{\hat{\pi}},\sys{\hat{H}}^*]+\hG^2\text{sinh}\bigg (\bom\bigg )\sum_{n=0}^{\infty}(n+1)(n+2)e^{-\beta\omt(n+\frac{1}{2})} \\
\nonumber &=\hD^2\text{Var}[\sys{\hat{\pi}},\sys{\hat{H}}^*]+2\hG^2\text{sinh}\bigg (\bom\bigg )\bigg(\frac{e^{-\bom}}{(1-e^{-\beta\omt})^3}\bigg), \\
&=\frac{\hD^2\omt^2}{4 \ \text{sinh}^{2}\big(\bom\big)}+2 \hG^2 \ \text{sinh}\bigg (\bom\bigg )\bigg(\frac{e^{-\bom}}{(1-e^{-\beta\omt})^3}\bigg),
\end{align}
where we used the series $\sum_{n=0}^{\infty} (n+1)(n+2)x^n=2/(1-x)^3$ for $|x|<1$ and that the variance of $\hmf$ is
\begin{align}
\text{Var}[\sys{\hat{\pi}},\sys{\hat{H}}^*]=\frac{\omt^2}{4 \ \text{sinh}^{2}\big(\bom\big)}.
\end{align}
We now compute the QFI, using~\eqref{eq:fishexp} and the fact that $p_n \pm p_{n+2}=p_n(1\pm e^{-2\beta\omt})$:
\begin{align}
\nonumber T^4\sys{F}(T)&=\sum_{n=0}^{\infty}p_nE_{nn}+4\sum_{n<m}\frac{(p_n-p_m)^2}{(p_n+p_m)\ln^2(\frac{p_n}{p_m})}E_{nm}, \\
\nonumber &=\hD^2\text{Var}[\sys{\hat{\pi}},\sys{\hat{H}}^*]+\frac{\hG^2}{\beta^2\omt^2}\sum_{n=0}^{\infty}\frac{(p_n-p_{n+2})^2}{p_n+p_{n+2}}(n+2)(n+1), \\
\nonumber &=\hD^2\text{Var}[\sys{\hat{\pi}},\sys{\hat{H}}^*]+\frac{\hG^2(1- e^{-2\beta\omt})^2}{\beta^2\omt^2(1+ e^{-2\beta\omt})}\sum_{n=0}^{\infty}p_n(n+2)(n+1), \\
\nonumber &=\hD^2\text{Var}[\sys{\hat{\pi}},\sys{\hat{H}}^*]+ \frac{\hG^2}{\beta^2\omt^2} \ \text{sinh}\bigg (\bom\bigg )\bigg(\frac{(1- e^{-2\beta\omt})^2}{(1+ e^{-2\beta\omt})}\bigg)\bigg(\frac{e^{-\bom}}{(1-e^{-\beta\omt})^3}\bigg), \\
&=\frac{\hD^2\omt^2}{4 \ \text{sinh}^{2}\big(\bom\big)}+\frac{\hG^2}{\beta^2\omt^2} \ \text{sinh}\bigg (\bom\bigg )\bigg(\frac{(1- e^{-2\beta\omt})^2}{(1+ e^{-2\beta\omt})}\bigg)\bigg(\frac{e^{-\bom}}{(1-e^{-\beta\omt})^3}\bigg),
\end{align}
We also require the average skew information, which can be obtained using~\eqref{eq:part2}:
\begin{align}
\nonumber  Q[\sys{\hat{\pi}},\enmf]&=\sum_{n<m}\bigg(p_n+p_m-\frac{2(p_n-p_m)}{\ln p_n-\ln p_m} \bigg)E_{nm}, \\
\nonumber &=\frac{\hG^2}{4}\sum_{n=0}^{\infty}\bigg(p_n+p_{n+2}-\frac{2(p_n-p_{n+2})}{\ln p_n-\ln p_{n+2}}\bigg)(n+2)(n+1), \\
&=\hG^2\ \text{sinh}\bigg (\bom\bigg )\frac{e^{-\bom}}{(1-e^{-\beta\omt})^3}\bigg(1+e^{-2\beta\omt}-\frac{T}{\omt}(1-e^{-2\beta\omt})\bigg).
\end{align}
To calculate the term $\langle\partial_{\scriptscriptstyle T} \hat{E}^*\rangle$ we use~\eqref{eq:heatcap} from the main text:
\begin{align}
\nonumber \langle\partial_{\scriptscriptstyle T} \enmf\rangle&=\sys{C}(T)+\frac{Q[\sys{\hat{\pi}},\enmf]-\text{Var}[\sys{\hat{\pi}},\enmf]}{T^2}, \\
&=\sys{C}(T)-\frac{\hG^2}{T\omt}\text{sinh}\bigg (\bom\bigg )\bigg(\frac{(1-e^{-2\beta\omt})}{(1-e^{-\beta\omt})^3}\bigg)e^{-\bom}.
\end{align}

\end{document}